\documentclass{jaa}
\usepackage{natbib}
\usepackage{hyperref}
\usepackage{graphicx}
\def\apjs{{\it Astrophys.~J.~Suppl.}}

\def\apjl{{\it Astrophys.~J.~Lett.}}

\def\aap{{\it Astron.~Astrophys.}}

\begin{document}\sloppy

%%paper title
%%For line breaks \\ can be used within title
%\title{The Kinematics and Structure of Serpens\\}
\title{The enhanced YSO population in Serpens\\}

%%author names are separated by comma (,)
%%use \and before the last author name
%%use a * along with the number separated by comma
%% for the  author for correspondence
%%\textsuperscript{number} is used for affiliation
%%\affilOne, \affilTwo etc., upto \affilTwentyfive is possible
%%Please note the first letter after \affil is capitalised in the command
%%

\author{Priya Hasan\textsuperscript{1}, Mudasir Raja\textsuperscript{1}, Md Saifuddin\textsuperscript{1} and S N Hasan\textsuperscript{2}}
\affilOne{\textsuperscript{1} Department of Physics, Maulana Azad National Urdu University, Hyderabad 500032, India.\\
{2} Department of Mathematics, Maulana Azad National Urdu University, Hyderabad 500032, India.\\}
%\affilTwo{\textsuperscript{2}Department of Q, University Z, Place Pincode, Country.}

%%escape two column mode for title, affiliation and abstract
%%by giving \twocolumn command as shown

\twocolumn[{

\maketitle

%%include \corres to print the corresponding author Email id
\corres{priya.hasan@gmail.com}

%%include \msinfo for
%%manuscript information such as
%%received, revised and accepted dates
%%
\msinfo{1 October 2022}{1 October 2022}

%%abstract
\begin{abstract}
The Serpens Molecular Cloud is one of the most active sites of ongoing star formation at a distance of about 300 pc, and hence is very  well-suited for studies of young low-mass stars and sub-stellar objects.
In this paper, for the Serpens star forming region, we find potential members of the Young Stellar Objects population from the Gaia DR3 data and study their kinematics and  distribution. We compile a catalog of 656 YSOs from  available catalogs ranging from X-ray to the infrared. We use this as a reference set  and cross-match it to find 87 Gaia DR3 member stars to produce a control sample with revised parameters. We queried the DR3 catalog with these parameters and found 1196 stars. We then applied three different density-based machine learning algorithms (DBSCAN, OPTICS and
HDBSCAN) to this sample and found potential YSOs. The three clustering algorithms identified a common set of  822 YSO members from Gaia DR3 in this region.  We 
also classified these objects using 2MASS and WISE data to study their distribution and the progress of star formation in Serpens. 
\end{abstract}

%%insert keywords separated by 3 hyphens using \keywords{words}

}]
%%close the twocolumn escape here

%%include \doinum{number}for the DOI number in the header
%%include \volnum{number} for the volume number in the header
%%include \year{yyyy} for  year of publication in the header
%%include \pgrange{num--num} page range of article in the header
%%include \artcitid{num} for the article citation id
%%include \lp to print last page of the article
%%include \setcounter{page}{pagenum} for the exact starting page of the article

\doinum{12.3456/s78910-011-012-3}
\artcitid{\#\#\#\#}
\volnum{000}
\year{0000}
\pgrange{1--}
\setcounter{page}{1}
\lp{1}

\section{Introduction}

Star forming regions (SFRs) house embedded star clusters and are the birthplaces of stars which  provide the missing links in understanding the star formation (SF) process \citep{Ascenso_2017}. As these young clusters are embedded in gas and dust, optical techniques (like multi-color optical photometry or spectroscopy) are inefficient in identification of Young Stellar Objects (YSOs). Infrared (IR) data is well suited for observations of embedded clusters. Complementary data ranging from X-ray to millimeter wavelengths, and spectroscopic follow-ups of the newly discovered population of young stars in star forming regions  enrich our understanding of SF in these regions. It is difficult to identify the members of any SFR, especially for nearby regions (within 500 pc), because they occupy large areas of the projected sky and would take a substantial amount of observational time. This paper presents an updated sample of young stellar members of Serpens based on Gaia DR3 data using machine learning clustering techniques \citep{2019A&A...626A..80C}.

Serpens is an interesting star-forming region for which unbiased datasets exist \citep{Harvey_2007, 2006A&A...458..789D, 2009ApJ...692..973E}.
It was identified as a  site of active star formation by \citep{1974ApJ...191..111S}, extends several degrees around
the young variable star  $VV\  Ser$ and forms part of the large local dark cloud complex
called the Aquila Rift, which has been extensively mapped in several molecular line
surveys \citep{da04000r, 1987ApJ...322..706D, 2001ApJ...547..792D}. It is well-suited for studies of very young low-mass stars and sub-stellar objects because of its proximity  of 260 pc \citep{Harvey_2007} and young age of 1-5 Myr \citep{https://doi.org/10.48550/arxiv.0809.3652}.

As part of the NOAO survey program `Towards a Complete Near-Infrared Spectroscopic and Imaging Survey of Giant Molecular Clouds' (PI: E. A. Lada), the Serpens
Molecular Cloud was observed with the Florida Multi-Object Imaging Near-Infrared Grism Observational Spectrometer (FLAMINGOS) at the Kitt Peak National Observatory 2.1 m telescope. In an earlier paper \citep{2012ASInC...4...29H}, used this data to study the YSO population and made important inferences about the SF processes in Serpens. The paper discussed the distribution of young embedded
sources using the Nearest Neighbor Method applied to a carefully selected
sample of near-infrared excess (NIRX) stars that trace star formation in the complex and identified  six
clusters, of which three were not earlier reported in literature. A
median age of 1-2 Myr and  a mean distance of 300 pc for the cluster was determined.

The Spitzer Legacy Survey ‘Molecular Cores to Planet Forming Disks’ Core to Disks (c2d) \citep{2009ApJS..181..321E} in Serpens shows evidence of sequential star formation from SW to NE in the main Serpens Core. The surface density of young stars in this region is much higher, by a factor of 10-100, than that of the other star-forming regions mapped by c2d  \citep{2009ApJS..181..321E}. It is an ideal region to build a ‘template’ for the study of disk evolution up to a few Myr within a well defined region by multi-wavelength observations of young stars and sub-stellar objects.

 \cite{2010ApJ...716..634G} made a spectroscopic study of the Serpens core. The Serpens Main Cluster, known since mid 70s, is made of two compact protoclusters, lying in a 0.6 pc long filamentary structure, along NW-SE. The two sub-clusters have similar masses within similar sized regions ($\approx 30 M\odot$ in 0.025 pc$^2$) each and an average
age of $10^5$ yr but differ in their velocity structures and molecular emission. The NW
cluster  devoid of bright NIR sources, has outflows powered by deeply embedded
Class 0 and I protostars.  \cite{2011A&A...528A..50D} inferred that star
formation was probably  triggered by the collision of two filament-like clouds. A large scale extinction map was presented by \cite{1999A&A...345..965C} . 

Unsupervised machine learning (ML) clustering techniques are used to find patterns or clusters in unlabeled databases. The problem of cluster recognition can be approached in a variety of ways using these methods, including centroid-based algorithms (like the $k$-means algorithm), distribution-based clustering (like Gaussian-mixture models), or density-based algorithms. (For an overview of clustering analysis in astronomy see  \cite{2012msma.book.....F}, Chap.~3.3 and references therein).

The density-based algorithms are particularly useful for locating clusters with arbitrary shapes that can be generically characterised as overdensities in a low density environment. They also have the benefit of not requiring any prior knowledge of the dataset being analysed. In other words, these algorithms do not assume any  distribution (such as one or many Gaussians) when associating the data points with a cluster, hence the user does not need to be aware of the number of clusters contained in the dataset. One of the most well-known methods in many fields is density-based spatial clustering of applications with noise (DBSCAN; \cite{Ester96adensity-based} and it is gaining popularity in astronomy  \citep{2018A&A...620A..27J, 2019A&A...626A..17C}. Ordering Points To Identify the Clustering Structure (OPTICS; \cite{inproceedings} and the hierarchical density-based spatial clustering of applications with noise (HDBSCAN; \cite{10.1007/978-3-642-37456-2_14} algorithms are improvements on DBSCAN that are gaining popularity due to their proven ability to detect different types of clusters.

Due to the young age of Serpens, we can assume that its members will have similar velocity distributions and will occupy a small area of the Galaxy. In contrast to the star population in the field, the cloud members should, in the multi-dimensional space described by their spatial coordinates and kinematic properties, appear to be grouped. In the five-dimensional space, which is defined by the three spatial coordinates and the two kinematic parameters proper motion in right ascension $\mu_{\alpha}^*$\footnote{$\mu_\alpha^*=\mu_\alpha cos\  \delta $ where $\mu_\alpha = \frac{\alpha_1-\alpha2}{\Delta t} $ is the apparent motion in right ascension in the time interval $\Delta \ t$ and $\delta$ is the declination}   and declination $\mu_\delta$, we ran the clustering algorithms. The DBSCAN, OPTICS, and HDBSCAN algorithms  utilised in this paper are from  \cite{scikit-learn}. By comparing their results, we aim to reduce the bias in selection that is inherent in each algorithm and provide a more reliable sample of YSO candidates for Serpens members. 

The paper is planned as follows: Section~1 is the introduction and the motivation for this work. Section~ 2 of our study provides a description of the data and sample construction we used. 
The three algorithms are applied to our Gaia sample in Section 3 where we also describe our methodology.  In Section 4, we go over the characteristics of this sample and  present the Two Micron All Sky Survey (2MASS) and Wide-field Infrared Survey Explorer (WISE) photometry and classification of our sample. Section~5 contains the Summary and Conclusions of our work. 

%\cite{roman08} presented the results obtained by NIR imaging of the Rosette complex using similar data.

\section{Data and Sample construction}
\subsection{Initial sample}
Gaia provides high-precision astrometric data (positions: right ascension ($\alpha$) and declination ($\delta$), parallax ($ \varpi $), and proper motions in right ascension ($\mu_\alpha$) and in declination ($\mu_\delta$)  which is of great significance to studies of open clusters \citep{2016, 2022arXiv220800211G}. 
\begin{figure}[h]
     \begin{center}
     \includegraphics[width=0.45\textwidth]{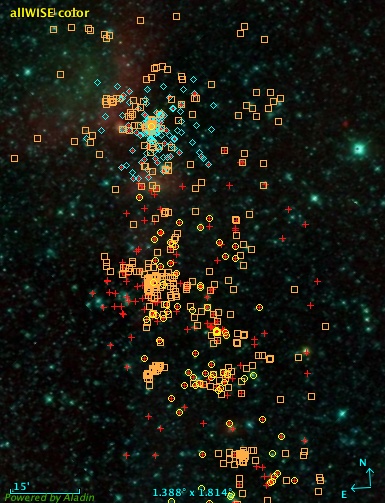}
     \caption{Data plotting on WISE image: Spitzer c2D \citep{Harvey_2007} (286) red plus
FIRX \cite{2012ASInC...4...29H}) (345) orange squares
Xray \cite{2009AJ....137.4777W} (138) blue rhombus,
R,Z \cite{2010A&A...513A..38S} (78) yellow circles
Total:656 unique sources }
\label{data}
\end{center}
\end{figure}

We began by compiling a list of YSOs in Serpens shown in Fig. \ref{data} and matching it with the 2MASS \cite{2006AJ....131.1163S} catalog. 
\begin{itemize}
    \item 
The Spitzer Legacy c2D Survey “Molecular Cores to Planet Forming Disks” included a 0.89 deg$^2$ area of Serpens. The  High Reliability Catalog included 377,456 total sources with
286 candidate YSOs \citep{Harvey_2007}.
\item 
\cite{2009AJ....137.4777W}  included a sample of 137 YSOs obtained from Chandra X-ray data in the Serpens core region. 
\item 
The Florida Multi-Object Imaging Near-Infrared Grism Observational Spectrometer (FLAMINGOS) described in 
\cite{2012ASInC...4...29H} includes a sample of 345 YSOs.
\item 
\cite{2009ApJ...691..672O} took 78 optical spectra in Serpens and found  58 stars (75\%) were  confirmed to be young, mostly K- and M-type stars that  belong to the cloud.
\item
\cite{2010A&A...513A..38S}
present a deep optical/near-infrared imaging survey of the Serpens molecular cloud as complementary optical data to the c2d Legacy survey  to study the star/disk formation and evolution in this cloud. 

\item
 \cite{2019ApJ...878..111H}
used Gaia DR2 parallaxes and proper motions to statistically measure $\approx$ 1167 kinematic members of Serpens, to evaluate the star formation history of the complex in a very large area of $\approx 36 \times 34$ degrees. We will compare our results with the above ones.
\end{itemize}

We combined the above catalogs (Fig \ref{data}) to obtain 656 unique sources, matched them first with 2MASS \cite{2006AJ....131.1163S} and then with Gaia DR3. This method is preferable to a sky cross-match by coordinates because it does not require to transform the 2MASS coordinates from the J2000 to the J2015.5 epoch. We found that 250 sources matched with Gaia DR3 sources, but only 87 matched with members from \cite{2019ApJ...878..111H}.
 
\begin{figure}[h]
\begin{center}    %\includegraphics[width=0.5\textwidth]{control.png}
\includegraphics[width=0.5\textwidth]{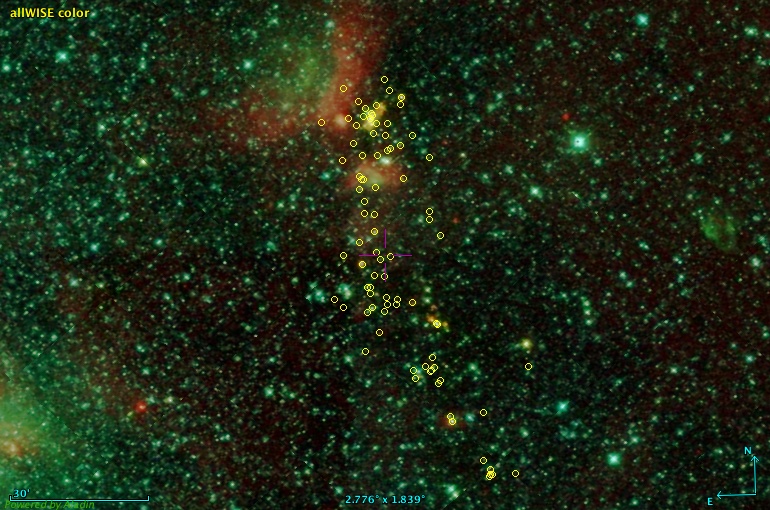}

%\caption{Distribution of parameters: distance, $\mu_\alpha$ and $\mu_\delta$ where the members  \cite{2019ApJ...878..111H} are in green (87), matches with DR3 in cyan (250) and the red histogram are the sources in the cutoffs matched sample of YSOs with}
\caption{WISE image of the Serpens
with RGB colours mapped to 22, 4.6, and 3.4 $\mu$m. The control sample members are represented as yellow circles.} 

\label{controlwise}
\end{center}
\end{figure}

For the 87 matched DR3 stars that are reliable members, we found the following average astrometric properties of the control sample listed in Table \ref{control} and shown in Fig. \ref{controlwise}.
Following \cite{Bailer_Jones_2015} we computed the individual distances as $d = 1/\varpi$ since the parallax fractional error of this
sample is lower than 10, and the average distance is 436.7 pc.

\begin{table}[h]
 \begin{tabular}{ |l|l|l|l|l|l|}
 \hline
% \multicolumn{4}{|c|}{Number of galaxies} \\
 %\hline
   Stats & $\alpha$ &$\delta$ & $\varpi$ & $\mu_{\alpha}^*$ &$\mu_{\delta}$\\ 
   & (deg) & (deg) & (mas) & (mas/yr) & (mas/yr) \\ \hline
Mean   & 277.4 & 0.75 & 2.29 & 2.19& -8.47\\
Sigma & 0.16 & 0.4 & 0.26 & 1.04& 0.68\\
 \hline
\end{tabular}
\caption{Mean and standard deviation ($1\sigma$) of the control sample.}
\label{control}
\end{table}

%\begin{figure}[h]
%     \begin{center}
%     \includegraphics[width=0.5\textwidth]%{gaia_yso.jpg}
%     \caption{A WISE image with the overlaid  matched stars with \cite{2019ApJ...878..111H} is in yellow (87), match with DR3 and the 1187 stars that were selected based on the criteria derived from the control sample}
%\label{gaia}
%\end{center}
%\end{figure}
%Figure \ref{controlhist} shows the distribution of the parameters.

We used these values to query the DR3 data in a $2^0$  radius from Serpens core, which is the most active region, with the following constraints: RUWE $<$ 1.4, RPlx $>$ 10, 
  5 $>$ $\mu_\alpha$ $>$ 1 
11.6 $<$ $\mu_\delta$ $<$ -6 
4.2  $>$ Plx $>$ 0 and derived  a sample of
1196 stars with an average value of 
$R_V$= -5.08 km/s for 66 stars, where $R_V$ is the radial velocity. 

%\subsubsection{Subsubsection heading.} Subsubsection text goes here (Radhakrishnan {\em et al.} 1980).

\section{Clustering Algorithms}
For our study, we considered the three spatial coordinates 
$$
X=d\  cos \ \delta\  cos\  \alpha$$
$$
Y=d\  cos\  \delta\  sin\  \alpha$$
$$
Z=d\  sin\  \delta  
$$
where $d$ is the distance computed as the inverse of parallax
and the two kinematic parameters proper motions in right ascension
$\mu_\alpha^*$ and in
declination
$\mu_\delta$.
 Given the low fraction of objects with radial velocity  measurements in our Gaia sample, we restricted the kinematic analysis to  only $\mu_\alpha^*$ and $\mu_\delta$.

We then applied the three clustering algorithms DBSCAN, OPTICS and HDBSCAN to our 5 parameters described above. Clusters are localised and arbitrary shaped regions of an N-dimensional space with an excess of points per volume unit. The points that do not satisfy this condition are classified as noise. The two parameters $\epsilon$ and $mPts$ are used to describe the density threshold. A sphere of radius $\epsilon$ is drawn around each point. If a minimum of $mPts$ points are found in the $\epsilon$ radius of a point, it is called a core point. Points which lie in the $\epsilon$ radius of a core point but do not have  minimum  $mPts$ are called border points and points outside the $\epsilon$ radius which do not have  minimum  $mPts$ are noise points. 
\subsection{DBSCAN}
DBSCAN was first introduced by  \cite{Ester96adensity-based}. The algorithm  strongly depends on the input parameters  $\epsilon$  and $mPts$ and uses it to identify clusters. We varied the values of $\epsilon$  and $mPts$ and obtained the cluster points  and noise points described in Table \ref{comp}. We find that more than 91.9\% of the control sample stars are identified using DBSCAN with the parameters used.

\begin{table}[h]
 \begin{tabular}{ |l|l|c|c|}
 \hline
    $\epsilon$  &$mPts$ & No of stars  & Control stars (\%) \\ 
 & &  (core points) & identified \\
    \hline
 0.5 & 50 & 978  & 96.5\\
 1.0 & 50 & 822  & 91.9\\
 1.5 & 50 & 1099   & 91.9\\

 \hline
\end{tabular}
\caption{Explored hyperparameters and number of cluster elements identified by DBSCAN. The last column shows the percentage of control sample stars identified.}
\label{comp}
\end{table}

%Explored hyperparameters and number of cluster elements identified by each algorithm. We recovered 1087 cluster points with 109 noise points with an $\epsilon$ of 1.5 and $mPts$=30 (Fig. \ref{dbscan}. 

\subsection{OPTICS}
By definition, all clusters discovered by DBSCAN in a given dataset have about the same density. Furthermore, in clusters with significant density gradients, such as a cluster made of a very dense core surrounded by a low density "halo," this algorithm struggles to identify all of the members. The hierarchical clustering algorithm Ordering Points To Identify the Clustering Structure (OPTICS) \cite{inproceedings} creates clusters with strong density gradients by exploring a range of $\epsilon$.

 \begin{figure}[h]
\begin{center}    \includegraphics[width=0.45\textwidth]{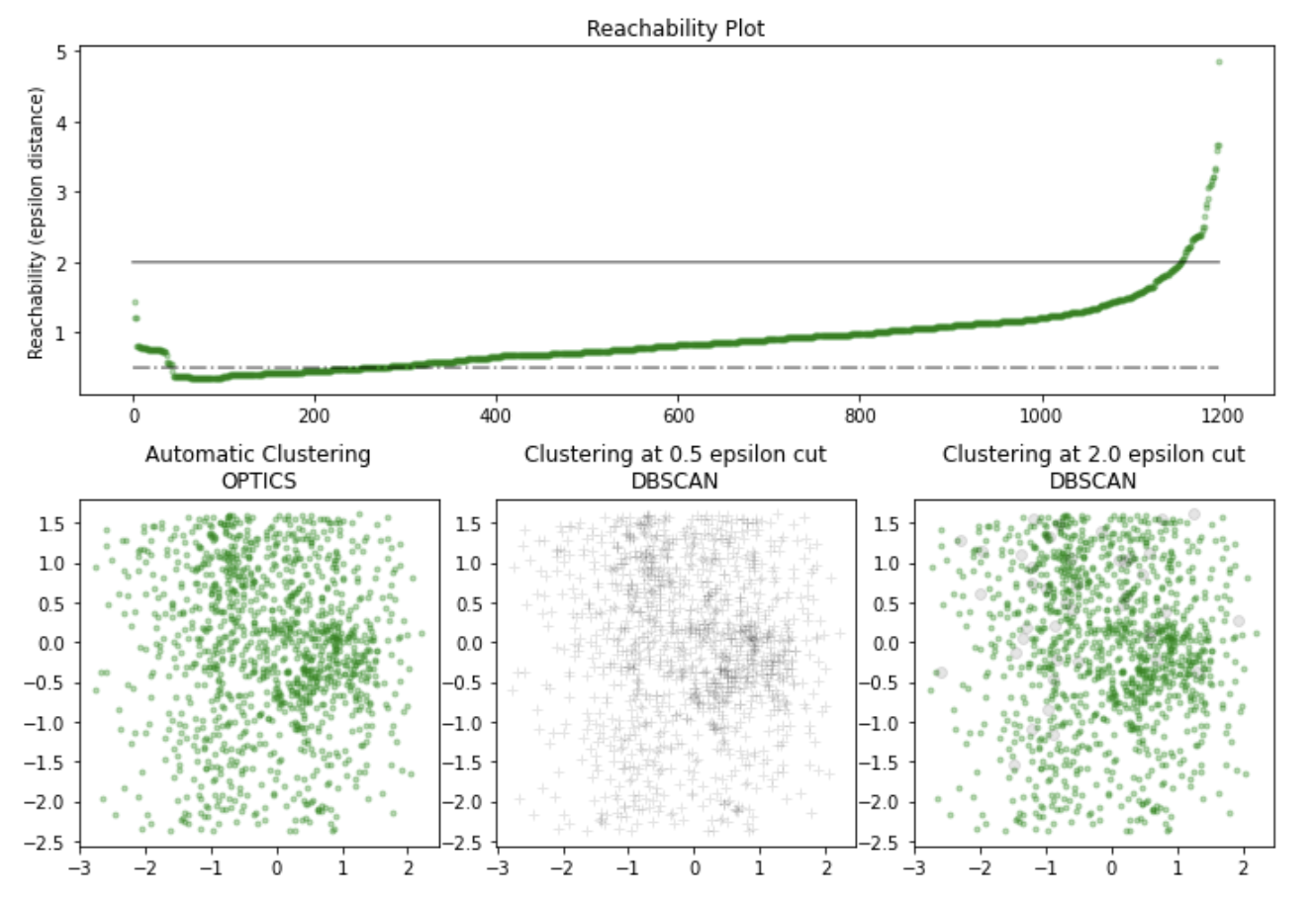}
     \caption{OPTICS: Reachability plot which shows the existance of only one cluster with $\epsilon \approx 0.5$
}
\label{optics}
\end{center}
\end{figure}

 OPTICS locates the cluster's densest areas and records this data in two variables called core distance and reachability distance. The former represents the distance between a core point and its nearest neighbour, and the latter is the maximum of the core distance of a core point . For a particular value of $mPts$, OPTICS organises the points into groups based on how far they can be reached from the densest region of the cluster. The reachability plot displays a string of distinctive troughs connected to individual potential clusters as a function of $\epsilon$. Figure \ref{optics} is the reachability plot for our sample and clearly shows a single valley with an $\epsilon$ close to 0.5.
 
\subsection{HDBSCAN}

Finding the ideal $\epsilon$ and $mPts$ values is challenging, which is a disadvantage of both DBSCAN and OPTICS. It is difficult to clearly identify the first and last points of the valleys in the reachability-distance plots produced by OPTICS and the step-like slope shift in the k-distance curves utilised by DBSCAN in high-density datasets. Finding suitable hyperparameters is made easier by the hierarchical method HBDSCAN because it only needs one hyperparameter $mCls$ (the "minimum cluster size") which is conceptually equivalent to $mPts$ \citep{10.1007/978-3-642-37456-2_14}. Similar to OPTICS, HDBSCAN is sensitive to the density gradients inside a cluster and can recognise clusters of various densities.

When we ran HDBSCAN, we found a cluster where the core points had 1103 stars. 
We matched these to the YSOs obtained by DBSCAN and OPTICS and got 822 common stars. These are stars which have a very high probability of being member YSOs.

\section{Infrared 2MASS and WISE photometry}
\begin{figure}[h]
     \begin{center}
     \includegraphics[width=0.45\textwidth]{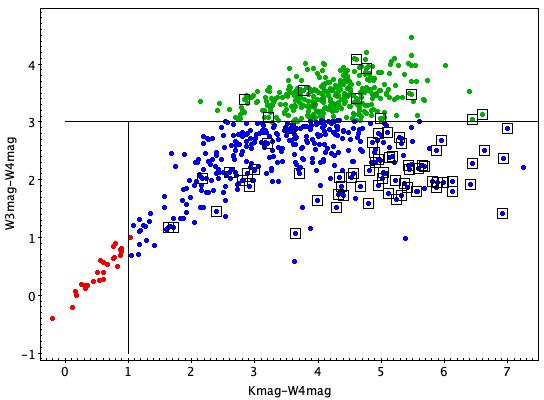}
     \caption{YSO Classification using 2MASS and WISE: The Class II stars are in green , Class III in blue and Photospheres in red. The stars from the control sample are the black squares.
}
\label{class}
\end{center}
\end{figure}

We cross-matched our data with 2MASS and WISE to 814 and 720 objects respectively. Figure \ref{class} shows the classification obtained for bare photospheres, Class II and Class III stars respectively using the method described in \cite{2012ApJ...744..130K}.

\begin{figure}[h]
     \begin{center}
     \includegraphics[width=0.45\textwidth]{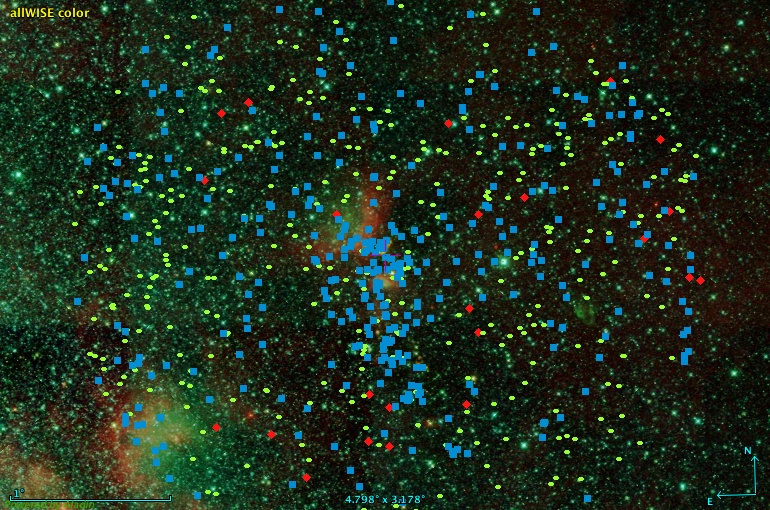}
     \caption{YSO Distribution using 2MASS and WISE colors, where Class II is green ovals, Class III is blue squares and photospheres is red rhombuses. 
}
\label{dist}
\end{center}
\end{figure}

We then plotted our YSOs on the WISE image to see the distribution of the sources (Fig. \ref{dist}.)

As found by \citep{2009ApJS..181..321E}  Serpens shows evidence of sequential star formation from SW to NE to the main Serpens Core. It was reported that SF has taken place along the South West to North East direction. In the figure, we see Class II and III stars seem distributed but   most of the bare photospheric stars are towards the west. Further studies are required to find ages of thse YSOs and trace the star formation. 
\section{Results and Conclusion}
This paper shows a unique method to identify young members of a star forming region, in this case, Serpens. YSOs are difficult to observe in the optical and hence other wavelengths ranging from Xray to IR are used in their identification and study. As Serpens is close to us, it occupies a very large region of the sky and  Gaia, being an all sky survey with unprecedented accuracy is  ideal to use for this purpose. 

In this work, we compiled YSO data of Serpens from various sources and wavelengths (656 stars) and matched it to Gaia DR3 data to find most probable YSO members (87 stars). This was used to build a control sample with data that was used to query Gaia DR3 to obtain  1196 stars. 

In the 5-parameter space of X, Y, Z and $\mu_\alpha^*$ and $\mu_\delta$ we 
 applied three different density-based machine learning algorithms (DBSCAN, OPTICS and
HDBSCAN) and found 822 common YSO members in the region. We found that they have similar values (due to our search criteria), but are spatially separated. We  classified these objects using 2MASS and WISE data  to find the distribution of Class II and Class III objects to study  their distribution. 
This is a potential method of increasing the YSO sample of star forming regions using machine learning techniques.  

\vspace{-2em}

%%Use section* for acknowledgements
\section*{Acknowledgements}
The authors would like to thank the referee for valuable comments that helped improve the paper.

This work has made use of data from the European Space Agency (ESA) mission
{\it Gaia} (\url{https://www.cosmos.esa.int/gaia}), processed by the {\it Gaia}
Data Processing and Analysis Consortium (DPAC,
\url{https://www.cosmos.esa.int/web/gaia/dpac/consortium}). Funding for the DPAC
has been provided by national institutions, in particular the institutions
participating in the {\it Gaia} Multilateral Agreement.
%Acknowledgements here.
\vspace{-1em}

%%use \balance somewhere in the left column of the last page to balance the two columns in the end page

%%References section
%\begin{theunbibliography}{}
\bibliographystyle{apj}
\bibliography{ref}

\begin{thebibliography}{}
\expandafter\ifx\csname natexlab\endcsname\relax\def\natexlab#1{#1}\fi

\bibitem[{Ankerst {$et~al$.}(1999)Ankerst, Breunig, Kriegel, \&
  Sander}]{inproceedings}
Ankerst, M., Breunig, M., Kriegel, H.-P., \& Sander, J. 1999, in OPTICS:
  Ordering Points to Identify the Clustering Structure, Vol.~28, 49--60

\bibitem[{Ascenso(2017)}]{Ascenso_2017}
Ascenso, J. 2017, in The Birth of Star Clusters (Springer International
  Publishing), 1--37

\bibitem[{Bailer-Jones(2015)}]{Bailer_Jones_2015}
Bailer-Jones, C. A.~L. 2015, Publications of the Astronomical Society of the
  Pacific, 127, 994

\bibitem[{{Cambr{\'e}sy}(1999)}]{1999A&A...345..965C}
{Cambr{\'e}sy}, L. 1999, Astronomy \& Astrophysics, 345, 965

\bibitem[{Campello {$et~al$.}(2013)Campello, Moulavi, \&
  Sander}]{10.1007/978-3-642-37456-2_14}
Campello, R. J. G.~B., Moulavi, D., \& Sander, J. 2013, in Advances in
  Knowledge Discovery and Data Mining, ed. J.~Pei, V.~S. Tseng, L.~Cao,
  H.~Motoda, \& G.~Xu (Berlin, Heidelberg: Springer Berlin Heidelberg),
  160--172

\bibitem[{{C{\'a}novas} {$et~al$.}(2019){C{\'a}novas}, {Cantero}, {Cieza},
  {Bombrun}, {Lammers}, {Mer{\'\i}n}, {Mora}, {Ribas}, \&
  {Ru{\'\i}z-Rodr{\'\i}guez}}]{2019A&A...626A..80C}
{C{\'a}novas}, H., {Cantero}, C., {Cieza}, L., {$et~al$.} 2019, \aap, 626, A80

\bibitem[{{Cantat-Gaudin} {$et~al$.}(2019){Cantat-Gaudin}, {Jordi}, {Wright},
  {Armstrong}, {Vallenari}, {Balaguer-N{\'u}{\~n}ez}, {Ramos}, {Bossini},
  {Padoan}, {Pelkonen}, {Mapelli}, \& {Jeffries}}]{2019A&A...626A..17C}
{Cantat-Gaudin}, T., {Jordi}, C., {Wright}, N.~J., {$et~al$.} 2019, \aap, 626,
  A17

\bibitem[{{Dame} {$et~al$.}(2001){Dame}, {Hartmann}, \&
  {Thaddeus}}]{2001ApJ...547..792D}
{Dame}, T.~M., {Hartmann}, D., \& {Thaddeus}, P. 2001, Astrophysics Journal,
  547, 792

\bibitem[{Dame \& Thaddeus(1985)}]{da04000r}
Dame, T.~M., \& Thaddeus, P. 1985, Astrophys. J., 297, 751

\bibitem[{{Dame} {$et~al$.}(1987){Dame}, {Ungerechts}, {Cohen}, {de Geus},
  {Grenier}, {May}, {Murphy}, {Nyman}, \& {Thaddeus}}]{1987ApJ...322..706D}
{Dame}, T.~M., {Ungerechts}, H., {Cohen}, R.~S., {$et~al$.} 1987, Astrophys.
  J., 322, 706

\bibitem[{{Djupvik} {$et~al$.}(2006){Djupvik}, {Andr{\'e}}, {Bontemps},
  {Motte}, {Olofsson}, {G{\r{a}}lfalk}, \& {Flor{\'e}n}}]{2006A&A...458..789D}
{Djupvik}, A.~A., {Andr{\'e}}, P., {Bontemps}, S., {$et~al$.} 2006, Astronomy
  \& Astrophysics, 458, 789

\bibitem[{{Duarte-Cabral} {$et~al$.}(2011){Duarte-Cabral}, {Dobbs}, {Peretto},
  \& {Fuller}}]{2011A&A...528A..50D}
{Duarte-Cabral}, A., {Dobbs}, C.~L., {Peretto}, N., \& {Fuller}, G.~A. 2011,
  \aap, 528, A50

\bibitem[{Eiroa {$et~al$.}(2008)Eiroa, Djupvik, \&
  Casali}]{https://doi.org/10.48550/arxiv.0809.3652}
Eiroa, C., Djupvik, A.~A., \& Casali, M.~M. 2008, The Serpens Molecular Cloud,
  doi:10.48550/ARXIV.0809.3652

\bibitem[{{Enoch} {$et~al$.}(2009){Enoch}, {Evans}, {Sargent}, \&
  {Glenn}}]{2009ApJ...692..973E}
{Enoch}, M.~L., {Evans}, Neal~J., I., {Sargent}, A.~I., \& {Glenn}, J. 2009,
  \apj, 692, 973

\bibitem[{Ester {$et~al$.}(1996)Ester, Kriegel, Sander, \&
  Xu}]{Ester96adensity-based}
Ester, M., Kriegel, H.-P., Sander, J., \& Xu, X. 1996, in KDD-96 Proceedings
  (AAAI Press), 226--231

\bibitem[{{Evans} {$et~al$.}(2009){Evans}, {Dunham}, {J{\o}rgensen}, {Enoch},
  {Mer{\'\i}n}, {van Dishoeck}, {Alcal{\'a}}, {Myers}, {Stapelfeldt}, {Huard},
  {Allen}, {Harvey}, {van Kempen}, {Blake}, {Koerner}, {Mundy}, {Padgett}, \&
  {Sargent}}]{2009ApJS..181..321E}
{Evans}, Neal~J., I., {Dunham}, M.~M., {J{\o}rgensen}, J.~K., {$et~al$.} 2009,
  Astrophysical Journal Supplement Series, 181, 321

\bibitem[{Feigelson \& Babu(2012)}]{2012msma.book.....F}
Feigelson, E.~D., \& Babu, G.~J. 2012, Modern Statistical Methods for
  Astronomy: With R Applications (Cambridge University Press),
  doi:10.1017/CBO9781139015653

\bibitem[{{Gaia Collaboration} {$et~al$.}(2022){Gaia Collaboration},
  {Vallenari}, {Brown}, {Prusti}, {de Bruijne}, {Arenou}, {Babusiaux},
  {Biermann}, {Creevey}, {Ducourant}, {Evans}, {Eyer}, {Guerra}, {Hutton},
  {Jordi}, {Klioner}, {Lammers}, {Lindegren}, {Luri}, {Mignard}, {Panem},
  {Pourbaix}, {Randich}, {Sartoretti}, {Soubiran}, {Tanga}, {Walton},
  {Bailer-Jones}, {Bastian}, {Drimmel}, {Jansen}, {Katz}, {Lattanzi}, {van
  Leeuwen}, {Bakker}, {Cacciari}, {Casta{\~n}eda}, {De Angeli}, {Fabricius},
  {Fouesneau}, {Fr{\'e}mat}, {Galluccio}, {Guerrier}, {Heiter}, {Masana},
  {Messineo}, {Mowlavi}, {Nicolas}, {Nienartowicz}, {Pailler}, {Panuzzo},
  {Riclet}, {Roux}, {Seabroke}, {Sordo{\o}rcit}, {Th{\'e}venin},
  {Gracia-Abril}, {Portell}, {Teyssier}, {Altmann}, {Andrae}, {Audard},
  {Bellas-Velidis}, {Benson}, {Berthier}, {Blomme}, {Burgess}, {Busonero},
  {Busso}, {C{\'a}novas}, {Carry}, {Cellino}, {Cheek}, {Clementini},
  {Damerdji}, {Davidson}, {de Teodoro}, {Nu{\~n}ez Campos}, {Delchambre},
  {Dell'Oro}, {Esquej}, {Fern{\'a}ndez-Hern{\'a}ndez}, {Fraile}, {Garabato},
  {Garc{\'\i}a-Lario}, {Gosset}, {Haigron}, {Halbwachs}, {Hambly}, {Harrison},
  {Hern{\'a}ndez}, {Hestroffer}, {Hodgkin}, {Holl}, {Jan{\ss}en}, {Jevardat de
  Fombelle}, {Jordan}, {Krone-Martins}, {Lanzafame}, {L{\"o}ffler}, {Marchal},
  {Marrese}, {Moitinho}, {Muinonen}, {Osborne}, {Pancino}, {Pauwels},
  {Recio-Blanco}, {Reyl{\'e}}, {Riello}, {Rimoldini}, {Roegiers}, {Rybizki},
  {Sarro}, {Siopis}, {Smith}, {Sozzetti}, {Utrilla}, {van Leeuwen}, {Abbas},
  {{\'A}brah{\'a}m}, {Abreu Aramburu}, {Aerts}, {Aguado}, {Ajaj},
  {Aldea-Montero}, {Altavilla}, {{\'A}lvarez}, {Alves}, {Anders}, {Anderson},
  {Anglada Varela}, {Antoja}, {Baines}, {Baker}, {Balaguer-N{\'u}{\~n}ez},
  {Balbinot}, {Balog}, {Barache}, {Barbato}, {Barros}, {Barstow},
  {Bartolom{\'e}}, {Bassilana}, {Bauchet}, {Becciani}, {Bellazzini},
  {Berihuete}, {Bernet}, {Bertone}, {Bianchi}, {Binnenfeld}, {Blanco-Cuaresma},
  {Blazere}, {Boch}, {Bombrun}, {Bossini}, {Bouquillon}, {Bragaglia},
  {Bramante}, {Breedt}, {Bressan}, {Brouillet}, {Brugaletta}, {Bucciarelli},
  {Burlacu}, {Butkevich}, {Buzzi}, {Caffau}, {Cancelliere}, {Cantat-Gaudin},
  {Carballo}, {Carlucci}, {Carnerero}, {Carrasco}, {Casamiquela}, {Castellani},
  {Castro-Ginard}, {Chaoul}, {Charlot}, {Chemin}, {Chiaramida}, {Chiavassa},
  {Chornay}, {Comoretto}, {Contursi}, {Cooper}, {Cornez}, {Cowell}, {Crifo},
  {Cropper}, {Crosta}, {Crowley}, {Dafonte}, {Dapergolas}, {David}, {David},
  {de Laverny}, {De Luise}, {De March}, {De Ridder}, {de Souza}, {de Torres},
  {del Peloso}, {del Pozo}, {Delbo}, {Delgado}, {Delisle}, {Demouchy},
  {Dharmawardena}, {Di Matteo}, {Diakite}, {Diener}, {Distefano}, {Dolding},
  {Edvardsson}, {Enke}, {Fabre}, {Fabrizio}, {Faigler}, {Fedorets}, {Fernique},
  {Fienga}, {Figueras}, {Fournier}, {Fouron}, {Fragkoudi}, {Gai},
  {Garcia-Gutierrez}, {Garcia-Reinaldos}, {Garc{\'\i}a-Torres}, {Garofalo},
  {Gavel}, {Gavras}, {Gerlach}, {Geyer}, {Giacobbe}, {Gilmore}, {Girona},
  {Giuffrida}, {Gomel}, {Gomez}, {Gonz{\'a}lez-N{\'u}{\~n}ez},
  {Gonz{\'a}lez-Santamar{\'\i}a}, {Gonz{\'a}lez-Vidal}, {Granvik}, {Guillout},
  {Guiraud}, {Guti{\'e}rrez-S{\'a}nchez}, {Guy}, {Hatzidimitriou}, {Hauser},
  {Haywood}, {Helmer}, {Helmi}, {Sarmiento}, {Hidalgo}, {Hilger},
  {H{\l}adczuk}, {Hobbs}, {Holland}, {Huckle}, {Jardine}, {Jasniewicz},
  {Jean-Antoine Piccolo}, {Jim{\'e}nez-Arranz}, {Jorissen}, {Juaristi
  Campillo}, {Julbe}, {Karbevska}, {Kervella}, {Khanna}, {Kontizas},
  {Kordopatis}, {Korn}, {K{\'o}sp{\'a}l}, {Kostrzewa-Rutkowska},
  {Kruszy{\'n}ska}, {Kun}, {Laizeau}, {Lambert}, {Lanza}, {Lasne}, {Le
  Campion}, {Lebreton}, {Lebzelter}, {Leccia}, {Leclerc}, {Lecoeur-Taibi},
  {Liao}, {Licata}, {Lindstr{\o}m}, {Lister}, {Livanou}, {Lobel}, {Lorca},
  {Loup}, {Madrero Pardo}, {Magdaleno Romeo}, {Managau}, {Mann}, {Manteiga},
  {Marchant}, {Marconi}, {Marcos}, {Marcos Santos}, {Mar{\'\i}n Pina},
  {Marinoni}, {Marocco}, {Marshall}, {Polo}, {Mart{\'\i}n-Fleitas}, {Marton},
  {Mary}, {Masip}, {Massari}, {Mastrobuono-Battisti}, {Mazeh}, {McMillan},
  {Messina}, {Michalik}, {Millar}, {Mints}, {Molina}, {Molinaro}, {Moln{\'a}r},
  {Monari}, {Mongui{\'o}}, {Montegriffo}, {Montero}, {Mor}, {Mora},
  {Morbidelli}, {Morel}, {Morris}, {Muraveva}, {Murphy}, {Musella}, {Nagy},
  {Noval}, {Oca{\~n}a}, {Ogden}, {Ordenovic}, {Osinde}, {Pagani}, {Pagano},
  {Palaversa}, {Palicio}, {Pallas-Quintela}, {Panahi}, {Payne-Wardenaar},
  {Pe{\~n}alosa Esteller}, {Penttil{\"a}}, {Pichon}, {Piersimoni}, {Pineau},
  {Plachy}, {Plum}, {Poggio}, {Pr{\v{s}}a}, {Pulone}, {Racero}, {Ragaini},
  {Rainer}, {Raiteri}, {Rambaux}, {Ramos}, {Ramos-Lerate}, {Re Fiorentin},
  {Regibo}, {Richards}, {Rios Diaz}, {Ripepi}, {Riva}, {Rix}, {Rixon},
  {Robichon}, {Robin}, {Robin}, {Roelens}, {Rogues}, {Rohrbasser},
  {Romero-G{\'o}mez}, {Rowell}, {Royer}, {Ruz Mieres}, {Rybicki}, {Sadowski},
  {S{\'a}ez N{\'u}{\~n}ez}, {Sagrist{\`a} Sell{\'e}s}, {Sahlmann}, {Salguero},
  {Samaras}, {Sanchez Gimenez}, {Sanna}, {Santove{\~n}a}, {Sarasso},
  {Schultheis}, {Sciacca}, {Segol}, {Segovia}, {S{\'e}gransan}, {Semeux},
  {Shahaf}, {Siddiqui}, {Siebert}, {Siltala}, {Silvelo}, {Slezak}, {Slezak},
  {Smart}, {Snaith}, {Solano}, {Solitro}, {Souami}, {Souchay}, {Spagna},
  {Spina}, {Spoto}, {Steele}, {Steidelm{\"u}ller}, {Stephenson}, {S{\"u}veges},
  {Surdej}, {Szabados}, {Szegedi-Elek}, {Taris}, {Taylo}, {Teixeira},
  {Tolomei}, {Tonello}, {Torra}, {Torra}, {Torralba Elipe}, {Trabucchi},
  {Tsounis}, {Turon}, {Ulla}, {Unger}, {Vaillant}, {van Dillen}, {van Reeven},
  {Vanel}, {Vecchiato}, {Viala}, {Vicente}, {Voutsinas}, {Weiler}, {Wevers},
  {Wyrzykowski}, {Yoldas}, {Yvard}, {Zhao}, {Zorec}, {Zucker}, \&
  {Zwitter}}]{2022arXiv220800211G}
{Gaia Collaboration}, {Vallenari}, A., {Brown}, A.~G.~A., {$et~al$.} 2022,
  arXiv e-prints, arXiv:2208.00211

\bibitem[{{Gorlova} {$et~al$.}(2010){Gorlova}, {Steinhauer}, \&
  {Lada}}]{2010ApJ...716..634G}
{Gorlova}, N., {Steinhauer}, A., \& {Lada}, E. 2010, \apj, 716, 634

\bibitem[{Harvey {$et~al$.}(2007)Harvey, Merin, Huard, Rebull, Chapman, II, \&
  Myers}]{Harvey_2007}
Harvey, P., Merin, B., Huard, T.~L., {$et~al$.} 2007, The Astrophysical
  Journal, 663, 1149

\bibitem[{{Hasan}(2012)}]{2012ASInC...4...29H}
{Hasan}, P. 2012, in Astronomical Society of India Conference Series, Vol.~4,
  Astronomical Society of India Conference Series, 29

\bibitem[{{Herczeg} {$et~al$.}(2019){Herczeg}, {Kuhn}, {Zhou}, {Hatchell},
  {Manara}, {Johnstone}, {Dunham}, {Bhardwaj}, {Jose}, \&
  {Yuan}}]{2019ApJ...878..111H}
{Herczeg}, G.~J., {Kuhn}, M.~A., {Zhou}, X., {$et~al$.} 2019, Astrophys. J.,
  878, 111

\bibitem[{{Joncour} {$et~al$.}(2018){Joncour}, {Duch{\^e}ne}, {Moraux}, \&
  {Motte}}]{2018A&A...620A..27J}
{Joncour}, I., {Duch{\^e}ne}, G., {Moraux}, E., \& {Motte}, F. 2018, \aap, 620,
  A27

\bibitem[{{Koenig} {$et~al$.}(2012){Koenig}, {Leisawitz}, {Benford}, {Rebull},
  {Padgett}, \& {Assef}}]{2012ApJ...744..130K}
{Koenig}, X.~P., {Leisawitz}, D.~T., {Benford}, D.~J., {$et~al$.} 2012, \apj,
  744, 130

\bibitem[{{Oliveira} {$et~al$.}(2009){Oliveira}, {Mer{\'\i}n}, {Pontoppidan},
  {van Dishoeck}, {Overzier}, {Hern{\'a}ndez}, {Sicilia-Aguilar}, {Eiroa}, \&
  {Montesinos}}]{2009ApJ...691..672O}
{Oliveira}, I., {Mer{\'\i}n}, B., {Pontoppidan}, K.~M., {$et~al$.} 2009, \apj,
  691, 672

\bibitem[{Pedregosa {$et~al$.}(2011)Pedregosa, Varoquaux, Gramfort, Michel,
  Thirion, Grisel, Blondel, Prettenhofer, Weiss, Dubourg, Vanderplas, Passos,
  Cournapeau, Brucher, Perrot, \& Duchesnay}]{scikit-learn}
Pedregosa, F., Varoquaux, G., Gramfort, A., {$et~al$.} 2011, Journal of Machine
  Learning Research, 12, 2825

\bibitem[{Prusti {$et~al$.}(2016)Prusti, de~Bruijne, Brown, Vallenari,
  Babusiaux, Bailer-Jones, Bastian, Biermann, Evans, Eyer, Jansen, Jordi,
  Klioner, Lammers, Lindegren, Luri, Mignard, Milligan, Panem, Poinsignon,
  Pourbaix, Randich, Sarri, Sartoretti, Siddiqui, Soubiran, Valette, van
  Leeuwen, Walton, Aerts, Arenou, Cropper, Drimmel, H{\o}g, Katz, Lattanzi,
  O'Mullane, Grebel, Holland, Huc, Passot, Bramante, Cacciari, Casta{\~{n}
  }eda, Chaoul, Cheek, Angeli, Fabricius, Guerra, Hern{\'{a}}ndez,
  Jean-Antoine-Piccolo, Masana, Messineo, Mowlavi, Nienartowicz,
  Ord{\'{o}}{\~{n}}ez-Blanco, Panuzzo, Portell, Richards, Riello, Seabroke,
  Tanga, Th{\'{e}}venin, Torra, Els, Gracia-Abril, Comoretto, Garcia-Reinaldos,
  Lock, Mercier, Altmann, Andrae, Astraatmadja, Bellas-Velidis, Benson,
  Berthier, Blomme, Busso, Carry, Cellino, Clementini, Cowell, Creevey,
  Cuypers, Davidson, Ridder, de~Torres, Delchambre, Dell'Oro, Ducourant,
  Fr{\'{e}}mat, Garc{\'{\i}}a-Torres, Gosset, Halbwachs, Hambly, Harrison,
  Hauser, Hestroffer, Hodgkin, Huckle, Hutton, Jasniewicz, Jordan, Kontizas,
  Korn, Lanzafame, Manteiga, Moitinho, Muinonen, Osinde, Pancino, Pauwels,
  Petit, Recio-Blanco, Robin, Sarro, Siopis, Smith, Smith, Sozzetti, Thuillot,
  van Reeven, Viala, Abbas, Aramburu, Accart, Aguado, Allan, Allasia,
  Altavilla, {\'{A}}lvarez, Alves, Anderson, Andrei, Varela, Antiche, Antoja,
  Ant{\'{o}}n, Arcay, Atzei, Ayache, Bach, Baker, Balaguer-N{\'{u}}{\~{n}}ez,
  Barache, Barata, Barbier, Barblan, Baroni, y~Navascu{\'{e}}s, Barros,
  Barstow, Becciani, Bellazzini, Bellei, Garc{\'{\i}}a, Belokurov, Bendjoya,
  Berihuete, Bianchi, Bienaym{\'{e}}, Billebaud, Blagorodnova, Blanco-Cuaresma,
  Boch, Bombrun, Borrachero, Bouquillon, Bourda, Bouy, Bragaglia, Breddels,
  Brouillet, Brüsemeister, Bucciarelli, Budnik, Burgess, Burgon, Burlacu,
  Busonero, Buzzi, Caffau, Cambras, Campbell, Cancelliere, Cantat-Gaudin,
  Carlucci, Carrasco, Castellani, Charlot, Charnas, Charvet, Chassat,
  Chiavassa, Clotet, Cocozza, Collins, Collins, Costigan, Crifo, Cross, Crosta,
  Crowley, Dafonte, Damerdji, Dapergolas, David, David, Cat, de~Felice,
  de~Laverny, Luise, March, de~Martino, de~Souza, Debosscher, del Pozo, Delbo,
  Delgado, Delgado, di~Marco, Matteo, Diakite, Distefano, Dolding, Anjos,
  Drazinos, Dur{\'{a}}n, Dzigan, Ecale, Edvardsson, Enke, Erdmann, Escolar,
  Espina, Evans, Bontemps, Fabre, Fabrizio, Faigler, Falc{\~{a}}o, Casas, Faye,
  Federici, Fedorets, Fern{\'{a}}ndez-Hern{\'{a}}ndez, Fernique, Fienga,
  Figueras, Filippi, Findeisen, Fonti, Fouesneau, Fraile, Fraser, Fuchs,
  Furnell, Gai, Galleti, Galluccio, Garabato, Garc{\'{\i}}a-Sedano, Gar{\'{e}},
  Garofalo, Garralda, Gavras, Gerssen, Geyer, Gilmore, Girona, Giuffrida,
  Gomes, Gonz{\'{a}}lez-Marcos, Gonz{\'{a}}lez-N{\'{u}}{\~{n}}ez,
  Gonz{\'{a}}lez-Vidal, Granvik, Guerrier, Guillout, Guiraud, G{\'{u}}rpide,
  Guti{\'{e}}rrez-S{\'{a}}nchez, Guy, Haigron, Hatzidimitriou, Haywood, Heiter,
  Helmi, Hobbs, Hofmann, Holl, Holland, Hunt, Hypki, Icardi, Irwin,
  de~Fombelle, Jofr{\'{e}}, Jonker, Jorissen, Julbe, Karampelas, Kochoska,
  Kohley, Kolenberg, Kontizas, Koposov, Kordopatis, Koubsky, Kowalczyk,
  Krone-Martins, Kudryashova, Kull, Bachchan, Lacoste-Seris, Lanza, Lavigne,
  Poncin-Lafitte, Lebreton, Lebzelter, Leccia, Leclerc, Lecoeur-Taibi,
  Lemaitre, Lenhardt, Leroux, Liao, Licata, Lindstr{\o}m, Lister, Livanou,
  Lobel, Löffler, L{\'{o}}pez, Lopez-Lozano, Lorenz, Loureiro, MacDonald,
  Fernandes, Managau, Mann, Mantelet, Marchal, Marchant, Marconi, Marie,
  Marinoni, Marrese, Marschalk{\'{o}}, Marshall, Mart{\'{\i}}n-Fleitas,
  Martino, Mary, Matijevi{\v{c}}, Mazeh, McMillan, Messina, Mestre, Michalik,
  Millar, Miranda, Molina, Molinaro, Molinaro, Moln{\'{a}}r, Moniez,
  Montegriffo, Monteiro, Mor, Mora, Morbidelli, Morel, Morgenthaler, Morley,
  Morris, Mulone, Muraveva, Musella, Narbonne, Nelemans, Nicastro, Noval,
  Ord{\'{e}}novic, Ordieres-Mer{\'{e}}, Osborne, Pagani, Pagano, Pailler,
  Palacin, Palaversa, Parsons, Paulsen, Pecoraro, Pedrosa, Pentikäinen,
  Pereira, Pichon, Piersimoni, Pineau, Plachy, Plum, Poujoulet, Pr{\v{s}}a,
  Pulone, Ragaini, Rago, Rambaux, Ramos-Lerate, Ranalli, Rauw, Read, Regibo,
  Renk, Reyl{\'{e}}, Ribeiro, Rimoldini, Ripepi, Riva, Rixon, Roelens,
  Romero-G{\'{o}}mez, Rowell, Royer, Rudolph, Ruiz-Dern, Sadowski,
  Sell{\'{e}}s, Sahlmann, Salgado, Salguero, Sarasso, Savietto, Schnorhk,
  Schultheis, Sciacca, Segol, Segovia, Segransan, Serpell, Shih, Smareglia,
  Smart, Smith, Solano, Solitro, Sordo, Nieto, Souchay, Spagna, Spoto, Stampa,
  Steele, Steidelmüller, Stephenson, Stoev, Suess, Süveges, Surdej, Szabados,
  Szegedi-Elek, Tapiador, Taris, Tauran, Taylor, Teixeira, Terrett, Tingley,
  Trager, Turon, Ulla, Utrilla, Valentini, van Elteren, Hemelryck, van Leeuwen,
  Varadi, Vecchiato, Veljanoski, Via, Vicente, Vogt, Voss, Votruba, Voutsinas,
  Walmsley, Weiler, Weingrill, Werner, Wevers, Whitehead, Wyrzykowski, Yoldas,
  {\v{Z}}erjal, Zucker, Zurbach, Zwitter, Alecu, Allen, Prieto, Amorim,
  Anglada-Escud{\'{e}}, Arsenijevic, Azaz, Balm, Beck, Bernstein, Bigot,
  Bijaoui, Blasco, Bonfigli, Bono, Boudreault, Bressan, Brown, Brunet,
  Bunclark, Buonanno, Butkevich, Carret, Carrion, Chemin, Ch{\'{e}}reau,
  Corcione, Darmigny, de~Boer, de~Teodoro, de~Zeeuw, Luche, Domingues, Dubath,
  Fodor, Fr{\'{e}}zouls, Fries, Fustes, Fyfe, Gallardo, Gallegos, Gardiol,
  Gebran, Gomboc, G{\'{o}}mez, Grux, Gueguen, Heyrovsky, Hoar, Iannicola,
  Parache, Janotto, Joliet, Jonckheere, Keil, Kim, Klagyivik, Klar, Knude,
  Kochukhov, Kolka, Kos, Kutka, Lainey, LeBouquin, Liu, Loreggia, Makarov,
  Marseille, Martayan, Martinez-Rubi, Massart, Meynadier, Mignot, Munari,
  Nguyen, Nordlander, Ocvirk, O'Flaherty, Sanz, Ortiz, Osorio, Oszkiewicz,
  Ouzounis, Palmer, Park, Pasquato, Peltzer, Peralta, P{\'{e}}turaud,
  Pieniluoma, Pigozzi, Poels, Prat, Prod'homme, Raison, Rebordao, Risquez,
  Rocca-Volmerange, Rosen, Ruiz-Fuertes, Russo, Sembay, Vizcaino, Short,
  Siebert, Silva, Sinachopoulos, Slezak, Soffel, Sosnowska, Strai{\v{z}}ys, ter
  Linden, Terrell, Theil, Tiede, Troisi, Tsalmantza, Tur, Vaccari, Vachier,
  Valles, Hamme, Veltz, Virtanen, Wallut, Wichmann, Wilkinson, Ziaeepour, \&
  Zschocke}]{2016}
Prusti, T., de~Bruijne, J. H.~J., Brown, A. G.~A., {$et~al$.} 2016, Astronomy
  \& Astrophysics, 595, A1

\bibitem[{{Skrutskie} {$et~al$.}(2006){Skrutskie}, {Cutri}, {Stiening},
  {Weinberg}, {Schneider}, {Carpenter}, {Beichman}, {Capps}, {Chester},
  {Elias}, {Huchra}, {Liebert}, {Lonsdale}, {Monet}, {Price}, {Seitzer},
  {Jarrett}, {Kirkpatrick}, {Gizis}, {Howard}, {Evans}, {Fowler}, {Fullmer},
  {Hurt}, {Light}, {Kopan}, {Marsh}, {McCallon}, {Tam}, {Van Dyk}, \&
  {Wheelock}}]{2006AJ....131.1163S}
{Skrutskie}, M.~F., {Cutri}, R.~M., {Stiening}, R., {$et~al$.} 2006, \aj, 131,
  1163

\bibitem[{{Spezzi} {$et~al$.}(2010){Spezzi}, {Mer{\'\i}n}, {Oliveira}, {van
  Dishoeck}, \& {Brown}}]{2010A&A...513A..38S}
{Spezzi}, L., {Mer{\'\i}n}, B., {Oliveira}, I., {van Dishoeck}, E.~F., \&
  {Brown}, J.~M. 2010, Astronomy \& Astrophysics., 513, A38

\bibitem[{{Strom} {$et~al$.}(1974){Strom}, {Grasdalen}, \&
  {Strom}}]{1974ApJ...191..111S}
{Strom}, S.~E., {Grasdalen}, G.~L., \& {Strom}, K.~M. 1974, Astrophys. J., 191,
  111

\bibitem[{{Winston} {$et~al$.}(2009){Winston}, {Megeath}, {Wolk}, {Hernandez},
  {Gutermuth}, {Muzerolle}, {Hora}, {Covey}, {Allen}, {Spitzbart}, {Peterson},
  {Myers}, \& {Fazio}}]{2009AJ....137.4777W}
{Winston}, E., {Megeath}, S.~T., {Wolk}, S.~J., {$et~al$.} 2009, \aj, 137, 4777

\end{thebibliography}
\vspace{-1.5em}

\end{document}